\documentclass[aps,prd,twocolumn,floatfix,reprint,preprintnumbers,altaffilletter]{revtex4-1}
\usepackage{graphicx,url,amssymb,amsmath,longtable,rotating,color,units,wasysym,subfigure,epsfig,multirow,enumerate}
\usepackage[colorlinks,urlcolor=blue,citecolor=blue,linkcolor=blue]{hyperref}

\definecolor{darkgreen}{rgb}{0.01, 0.75, 0.24}

\begin{document}
\pacs{}

\title{Determining the population properties of spinning black holes}

\author{Colm Talbot}
\email{colm.talbot@monash.edu}
\affiliation{School of Physics and Astronomy, Monash University, Clayton, Victoria 3800, Australia}

\author{Eric Thrane}
\affiliation{School of Physics and Astronomy, Monash University, Clayton, Victoria 3800, Australia}
\affiliation{OzGrav: The ARC Centre of Excellence for Gravitational-wave Discovery, Hawthorn, Victoria 3122, Australia}

\date{\today}

\begin{abstract}
  There are at least two formation scenarios consistent with the first gravitational-wave observations of binary black hole mergers.
  In field models, black hole binaries are formed from stellar binaries that may undergo common envelope evolution.
  In dynamic models, black hole binaries are formed through capture events in globular clusters.
  Both classes of models are subject to significant theoretical uncertainties.
  Nonetheless, the conventional wisdom holds that the distribution of spin orientations of dynamically merging black holes is nearly isotropic while field-model black holes prefer to spin in alignment with the orbital angular momentum.
  We present a framework in which observations of black hole mergers can be used to measure ensemble properties of black hole spin such as the typical black hole spin misalignment.
  We show how to obtain constraints on population hyperparameters using minimal assumptions so that the results are not strongly dependent on the uncertain physics of formation models.
  These data-driven constraints will facilitate tests of theoretical models and help determine the formation history of binary black holes using information encoded in their observed spins.
  We demonstrate that the ensemble properties of binary detections can be used to search for and characterize the properties of two distinct populations of black hole mergers.
\end{abstract}

\maketitle

\section{Introduction}
  At present, merging black holes are the only directly detected source of gravitational waves~\cite{PhysRevLett.116.061102, PhysRevLett.116.241102, PhysRevLett.116.241103, PhysRevD.93.122003, PhysRevX.6.041014, PhysRevX.6.041015, GW150914_systematics, GW150914_applications, PhysRevLett.116.131103, GW170104}.
  A variety of mechanisms by which black hole binaries can form have been proposed.
  These mechanisms might yield significantly different distributions of the intrinsic parameters of binaries~\cite{Rodriguez2016}.
  In this work we focus on the distribution of spin orientations to probe black hole binary formation mechanisms.
  We consider two mechanisms which are expected to dominate, the field and dynamical models (see, e.g., \cite{0264-9381-27-11-114007} for a detailed review).

  In dynamical models, the binary forms when two black holes become gravitationally bound in dense stellar environments such as globular clusters \cite{doi:10.1093/mnras/173.3.729}.
  Due to mass segregation such clusters arrange themselves with more massive objects being found in the center and less massive objects on the outside.
  This means that binaries are expected to have mass ratios close to unity \cite{Sigurdsson1993}.
  It is expected that the spins of the two companions will be isotropically oriented \cite{Rodriguez2016}.

  The distribution of spin orientations in field models is subject to more theoretical uncertainty (e.g., \cite{Postnov2014}).
  In field models, a stellar binary forms and the components of the binary then coevolve.
  Although such stars are expected to form with their angular momenta aligned with the total angular momentum of the binary, there are exceptions (e.g., \cite{0004-637X-726-2-68, 0004-637X-785-2-83}).
  If binaries are formed with misaligned spins, tidal interactions and mass transfer processes between the stars can align the angular momenta of the stars with the total angular momentum of the binary (e.g., \cite{Bardeen1975, Hut1981}).
  When the first star explodes in a supernova and collapses to form a black hole, a natal kick may be imparted on the two companions due to asymmetry of the explosion (e.g., \cite{Janka2012}), increasing misalignment between spin and angular momentum vectors.
  The subsequent evolution of the secondary, possibly involving a common envelope phase, can reverse this misalignment \cite{Ivanova2013}.
  This is followed by the supernova of the secondary, which may give each black hole another kick and some additional degree of misalignment.
  The net effect is to leave the population of black hole spin orientations distributed about the angular momentum vector of the binary with some unknown typical misalignment angle~\cite{kalogera2000, Repetto2012, Rodriguez2016, OShaughnessy2017}.

  Following the formation of the black hole binary (either through dynamical capture or common evolution) the spin orientation of nonaligned spinning black holes changes due to  precession.
  Isotropic spin orientation distributions are expected to remain isotropic throughout such evolution~\cite{Bogdanovic2007}.
  However, anisotropic distributions, such as those predicted by field models, may change significantly~\cite{Schnittman2004, Kesden2015, Gerosa2015}.
  Here, we are interested in the distribution of spin orientations at the moment the binary enters LIGO's observing band.
  We therefore measure our spin orientations at $f_\text{ref}=\unit[20]{Hz}$.
  Advanced LIGO's observing band will eventually extend down to $\unit[10]{Hz}$, but we use $\unit[20]{Hz}$ here for the sake of convenience.
  One may use the spin orientation at $f_\text{ref}$ to reverse engineer the spin alignment distribution at the moment of formation, but this is not our present goal.

  In this paper, we use Bayesian hierarchical modeling (e.g., \cite{gelman2013bayesian}) and model selection to infer the parameters describing the distribution of spins of black hole binaries.
  We construct a mixture model, which treats the fraction of dynamical mergers, the fraction of isolated binary mergers, and the typical spin misalignment of the primary and secondary black holes as free parameters.
  We apply the model to simulated data (including noise) to show that we can both detect the presence of distinct populations, and also measure hyperparameters describing typical spin misalignment.

  Our method builds on a body of research using gravitational waves to study the ensemble properties of compact binaries.
  In \cite{Stevenson2015}, it was shown that Bayesian model selection can be used to distinguish between formation channels using nonparametrized mass distributions.
  Clustering was used in \cite{Mandel2017} to show that model-independent statements about the existence of distinct mass subpopulations can be made with an ensemble of detections.
  In \cite{Fishbach2017, Gerosa2017}, it was shown that the spin magnitude distribution can be used to determine whether observed merging black holes formed through hierarchical mergers of smaller black holes.
  Hierarchical merger models predict an isotropic distribution of black hole spin orientations since all binaries form through dynamical capture.

  Vitale \textit{et al.}~\cite{Vitale2017a} showed that model selection can be used to distinguish between models which predict mutually exclusive spin orientations of merging compact binaries, both binary black holes and neutron star black hole binaries.
  In order to generate two distinct populations with different spin distributions, binaries were generated with random spin angles.
  Those with tilt angles (between the black hole spin and the Newtonian orbital angular momentum) $<10^\circ$ were considered to be a fieldlike binary while those with tilt angles $>10^\circ$ were considered to be dynamiclike.
  The authors showed that, after $\sim100$ detections, one can recover the proportion of binaries in each population to within $\sim10\%$ at $1\sigma$.

  Stevenson \textit{et al.} \cite{Stevenson2017b} used Bayesian hierarchical modeling to recover the proportion of binaries taken from a set of four populations distributed according to astrophysically motivated, spin orientation distributions with fixed spin magnitudes ($a_i=0.7$).
  Unlike~\cite{Vitale2017a}, the populations overlap so that even precise knowledge of a binary's spin parameters does not provide certain knowledge about its parent population.
  Of the four populations, three are different distributions predicted by population synthesis models of isolated binary evolution and the fourth is the isotropic distribution predicted for dynamic formation.
  They achieve a similar result to Vitale \textit{et al.}, measuring the relative proportion of different populations at the $\sim10\%$ level after $100$ events.
  They also demonstrate that their two ``extreme hypotheses" (perfect alignment and isotropy) can be ruled out at $> 5\sigma$ after as few as five events if they are not good descriptions of nature.


  We build on these studies by employing a (hyper)parametrized model of the spin orientation distribution for the field model in order to measure not just the fraction of binaries from different populations, but also properties of the field model.
  In particular, we aim to {\em measure} the typical black hole misalignment for black hole binaries formed in the field.
  The advantage of this approach is that our modeling employs a broadly accepted idea from theoretical modeling (black holes in field binaries should be somewhat aligned) without assuming less certain details about the size of the misalignment.
  Since our model is agnostic with respect to the detailed physics of binary formation and subsequent evolution, the resulting methodology is robust against theoretical bias and provides a {\em measurement} of black hole spin misalignment for binaries formed in the field.

  The remainder of the paper is organized as follows.
  In the next section we review how the properties of merging binary black holes are recovered from observed data and briefly discuss the current observational results.
  We then introduce a useful parametrization to describe an admixture of field and dynamical black hole mergers.
  We follow this with a description hierarchical inference.
  We then present the results of a proof-of-principle study using simulated data.
  We introduce a new tool for visualizing spin orientations, spin maps.
  Finally, closing thoughts are provided.

\section{Gravitational-wave parameter estimation}\label{PE}
  In order to determine the parameters describing the sources of gravitational waves $\Theta$ from gravitational-wave strain data $h$, we employ Bayesian inference.
  Merging binary black hole waveforms are described by 15 parameters: two masses $\{m_1,m_2\}$, two three-dimensional spin vectors $\{\textbf{S}_1,\textbf{S}_2\}$,  and seven additional parameters to specify the position and orientation of the source relative to Earth.
  It is possible that in both the field and dynamical formation models the presence of a third companion will induce eccentricity when the binary enters LIGO’s observing band through Lidov-Kozai cycles~\cite{Antonini2017, Toonen2016, Lidov1962, Kozai1962, 0004-637X-598-1-419}.
  However, we consider only circular binaries.
  Most gravitational-wave parameter estimation results obtained to date have been obtained using the Bayesian parameter estimation code {\sc LALInference} \cite{PhysRevD.91.042003}.
  For our study we use the {\sc LALInference} implementation of nested sampling \cite{doi:10.1063/1.1835238}.
  We employ reduced order modeling and reduced order quadrature \cite{Smith2016} to  limit the computational time of the analysis.

  Performing parameter estimation over this 15-dimensional space is computationally intensive.
  In order to maximize the efficiency sampling this high-dimensional space, the effect of the two spin vectors on the waveform is approximately represented using two spin parameters \cite{Schmidt2015},
  \begin{equation}
  \begin{split}
  \chi_\text{eff} = & \frac{a_1\cos(\theta_1) + qa_2\cos(\theta_2)}{1+q} \\
  \chi_p = & \text{max}\left(a_1\sin(\theta_1), \left(\frac{4q+3}{4+3q}\right)qa_2\sin(\theta_2)\right).
  \end{split}
  \label{eq:chi}
  \end{equation}
  Here $(a_1, a_2)$ are the dimensionless spin magnitudes, $q=m_1 / m_2 <1$ is the mass ratio and $(\theta_1, \theta_2)$ are the angles between the spin angular momenta and the Newtonian orbital angular momentum of the binary.
  The variable $\chi_{\text{eff}}$ is ``the effective spin parameter."
  When $\chi_{\text{eff}} >0$, the binary merges at a higher frequency than for $\chi_{\text{eff}}=0$ and hence spends more time in the observing band~\cite{Campanelli2006}.
  Similarly, binaries with $\chi_{\text{eff}} <0$ spend less time in the observing band.
  The variable $\chi_p$ describes the precession of the binary, which is manifest as a long-period modulation of the signal~\cite{Apostolatos1994}.

  Using numerical relativity to compute all of the waveforms necessary for parameter estimation is computationally prohibitive.
  Parameter estimation therefore relies on ``approximants," which can be used for rapid waveform estimation.
  We use the {\sc IMRPhenomP} approximant~\cite{Hannam2014}, which has been used in many recent parameter estimation studies, including parameter estimation for recently observed binaries (e.g., \cite{PhysRevLett.116.241102,GW150914_systematics, GW170104}).
  {\sc IMRPhenomP} approximates a generically precessing binary waveform using $\chi_{\text{eff}}$ and $\chi_p$.
  Parameter estimation of the confirmed binary black hole detections, GW150914~\cite{PhysRevX.6.041014, PhysRevLett.116.241102}, GW151226~\cite{PhysRevLett.116.241103} and GW170104~\cite{GW170104}, yield (slightly) informative posterior distributions for $\chi_{\text{eff}}$.
  However, the posterior distributions for $\chi_p$ show no significant deviation from the prior.

  The observed distribution of these two effective spin parameters will depend on the mass and spin magnitude distributions of black holes.
  The distributions are expected to differ for binaries formed through different mechanisms \cite{kalogera2000}.
  We do not consider these effects.
  Instead we work directly with the spin orientations of each black hole.
  For our purposes, it will be useful to define two additional variables:
  \begin{align}
  z_1 = & \cos(\theta_1) \nonumber\\
  z_2 = & \cos(\theta_2) .
  \label{eq:z}
  \end{align}
  Instead of working with $\chi_\text{eff}$ and $\chi_p$, we work with distributions of $z_1, z_2$.
  We note that $z_i\approx1$ corresponds to aligned spin while $z_i\approx -1$ corresponds to antialigned spin and $z_i=0$ corresponds to black holes spinning in the orbital plane.

\section{Models}\label{models}
  For the purpose of this work we ignore the detailed formation history used in population synthesis studies.
  Instead, we introduce a simple parametrization designed to capture the salient features of the field and dynamic models.
  More sophisticated parametrizations are possible and will (eventually) be necessary to accurately describe realistic populations.
  However, we believe this is a suitable starting point given current theoretical uncertainty.

  We hypothesize that the distribution of $\{z_1, z_2\}$ can be approximated as an admixture of two populations.
  The first population is described by a truncated Gaussian peaked at $(z_1, z_2) = (1,1)$ with width $(\sigma_1, \sigma_2)$.
  This is our proxy for the population formed in the field.
  The Gaussian shape mimics the  form of distributions predicted by population synthesis models, which are clustered about $z=1$ with some unknown spread.
  The second population is uniform in $(z_1, z_2)$, this represents the dynamically formed population.
  The relative abundances of each population are given by $\xi$ (field) and $1 - \xi$ (dynamic).
  Thus, according to our parametrization, the true distribution of black hole mergers can be approximately described as follows:
  \begin{align}\label{eq:ps}
  p_0(z_1, z_2) & = \frac{1}{4} \nonumber\\
  p_1(z_1, z_2) & = \frac{2}{\pi}
  \frac{1}{\sigma_1} \frac{e^{-(z_1 - 1)^2/2\sigma_1^2}}
  {\text{erf}\left(\sqrt{2}/\sigma_1\right)}
  \frac{1}{\sigma_2} \frac{e^{-(z_2 - 1)^2/2\sigma_2^2}}{\text{erf}\left(\sqrt{2}/\sigma_2\right)} \\
  p(z_1,z_2) & = (1 - \xi) p_0 + \xi p_1
  \end{align}
  Here, $p_0(z_1, z_2)$ is the true dynamic-only distribution, $p_1(z_1, z_2)$ is the true field-only distribution, and $p(z_1, z_2)$ is the true distribution for all black hole binaries.
  These distributions depend on three hyperparameters: two widths $(\sigma_1, \sigma_2)$ and one fraction $\xi$.

\begin{figure}
  \includegraphics[width=\linewidth]{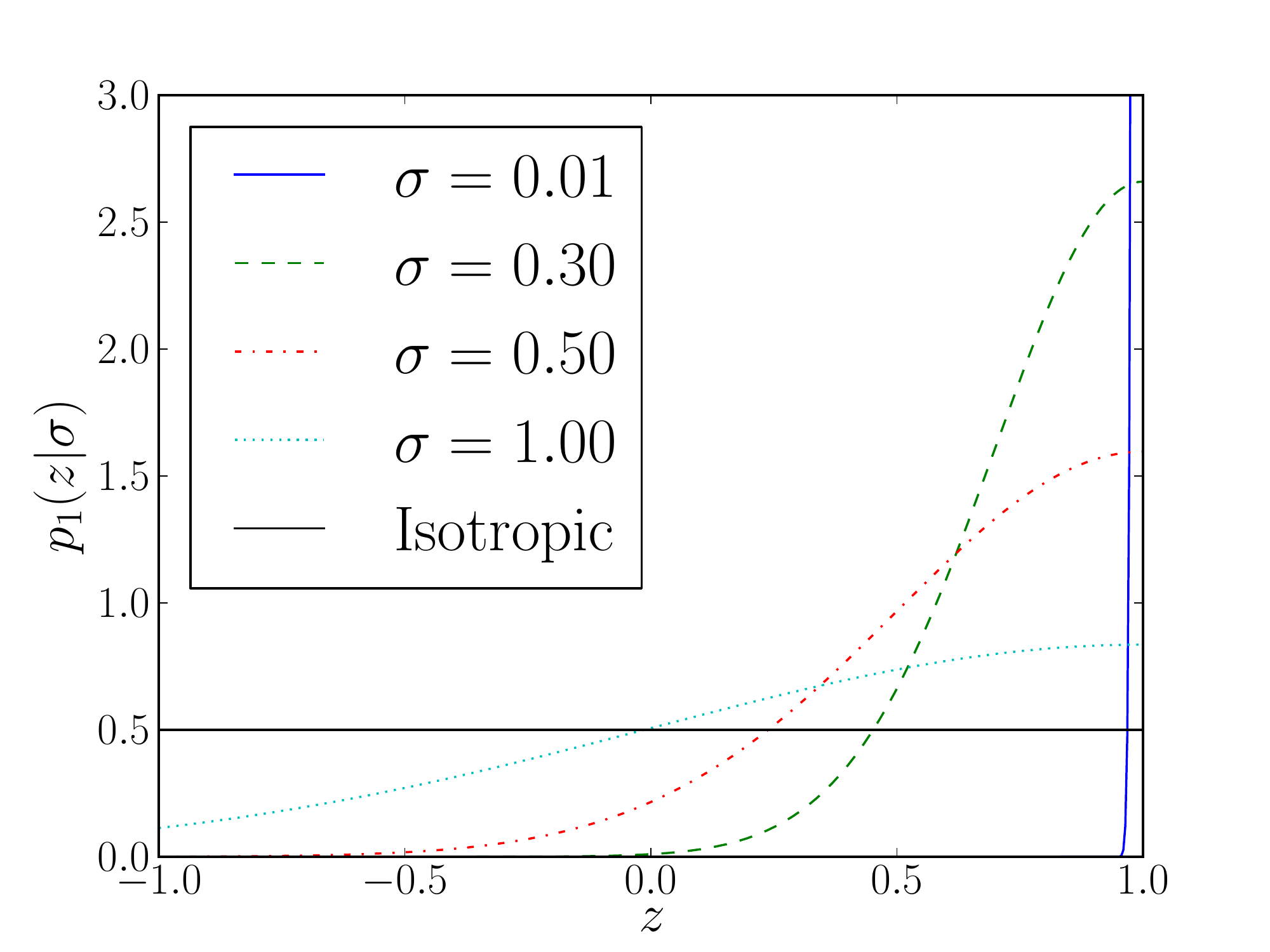}
  \caption{
  The distribution of $z$ for our field model proxy with varying $\sigma$; see Eq.~\ref{eq:ps}.
  By sending $\sigma\rightarrow0$, we obtain perfect alignment and by sending $\sigma\rightarrow\infty$, we obtain an isotropic distribution.
  }
  \label{fig:norm}
  \end{figure}

  For each of our population hyperparameters $\{\sigma_1, \sigma_2, \xi$\}, we choose uniform prior distributions between 0 and 1.
  For $\xi$ this covers the full allowed range of values.
  For $\sigma$, this prior is chosen to be consistent with the most conservative estimates on spin misalignments predicted by  field models (isotropically distributed kicks with the same velocity distribution as neutron stars, isotropic full kicks in \cite{Rodriguez2016}).
  In Fig.~\ref{fig:norm}, we plot $p_1$ for various values of $\sigma$.

  There are two interesting limiting cases.
  We note that $p_1(z|\sigma)\rightarrow\delta(z-1)$ as $\sigma\rightarrow 0$.
  This corresponds to perfect alignment of black hole spins.
  We also note that $p_1(z|\sigma) \rightarrow p_0$ as $\sigma \rightarrow \infty$.
  Thus, depending on the choice of prior, the dynamical model is degenerate with the field model evaluated at one point in hyperparameter space.
  A consequence of this limiting behavior is that it is far more difficult to distinguish samples drawn from a broad aligned distribution ($\sigma=1$), than an \textit{almost} perfectly aligned distribution ($\sigma=0.01$).
  It is simple to extend this model to include more terms describing additional subpopulations or alter the form of the existing terms to better fit physically motivated distributions.

\section{Bayesian hierarchical modeling}\label{hierarchical}
  Bayesian hierarchical modeling involves splitting a Bayesian inference problem into multiple stages.
  In the case of merging compact binaries these steps are as follows:
  \begin{enumerate}[i]
  \itemsep0em
  \item Perform gravitational-wave parameter estimation as described above.
        We adopt priors that are uniform in spin magnitude and isotropic in spin orientations.
  \item Assume the population from which events are drawn is described by hyperparameters $\Lambda$.
        Calculate a likelihood function for the data given $\Lambda$ by marginalizing over the parameters for individual events $\Theta$.
  \item Combine multiple events to derive a joint likelihood for $\Lambda$.
  \item Use the joint likelihood to derive posterior distributions for $\Lambda$, which, in turn, may be used to construct Bayes factors or odds ratios comparing different population models and confidence intervals on hyperparameters.
  \end{enumerate}

  Step~(i) produces a set of $n_k$ posterior samples $\{\Theta_i\}$, sampled according to the likelihood of the binary having each set of parameters, $p(\Theta|h)$.
  This step is computationally expensive and requires the application of a specialized tool such as {\sc LALInference}.
  In Step~(ii), we estimate $\Lambda$ using the posterior samples $\{z_i\}$.
  Our likelihood requires marginalization over $z$, for each event.
  Since {\sc LALInference} approximates the posterior for $\Theta$ with a list of posterior sample points, the marginalization integral over $(z_1, z_2)$ can be approximated by summing the probability of each sample in the {\sc LALInference} posterior chain for our population model (see, e.g., \cite[Chapter~29]{MacKay:2002:ITI:971143} for details).

  Step (iii): To combine data from $N$ events, we multiply the likelihoods:
  \begin{align}
  \mathcal{L}_k(h_k|\Lambda) & = \int dz_1 dz_2 \, p \left( z_1, z_2|h_k \right) p \left( z_1, z_2 | \Lambda \right) \nonumber\\
  & = \frac{1}{n_k} \sum_{\alpha = 1}^{n_k} p \left( z_{\alpha1}, z_{\alpha2} | \Lambda \right) \\
  \mathcal{L}(\{h_k\}|\Lambda) & = \prod_{k=1}^N \mathcal{L}_k(h_k|\Lambda).
  \label{eq:l}
  \end{align}
  Here, $\mathcal{L}_k(h_k|\Lambda)$ is the likelihood function for the $k$th event with strain data $h_k$.
  The joint likelihood function $\mathcal{L}(\{h_k\}|\Lambda)$ combines data from all $N$ measurements to arrive at the best possible constraints on $\Lambda$.

  Step (iv): At last, we arrive at the posterior distribution for $\Lambda$, $p(\Lambda|\{h_k\})$.
  Combining the joint likelihood $\mathcal{L}(\{h_k\}|\Lambda)$ with a prior distribution for the hyperparameters $\Lambda$, $\pi(\Lambda|H)$,
  for a particular population model, $H$, we obtain
  \begin{equation}
  \begin{split}
  p(\Lambda|\{h_k\}) & = \frac{\mathcal{L}(\{h_k\}|\Lambda)\pi(\Lambda|H)}{Z(\{h_k\}|H)} \\
  & = \frac{\pi(\Lambda|H)}{Z(\{h_k\}|H)} \prod_{k=1}^N \frac{1}{n_k} \sum_{\alpha = 1}^{n_k} p \left( z_{\alpha1}, z_{\alpha2}|\Lambda \right) \\
  & \propto \prod_{k=1}^N \sum_{\alpha = 1}^{n_k} p(z_{\alpha1}, z_{\alpha1} |\Lambda).
  \end{split}
  \label{eq:post}
  \end{equation}
  Here, $Z(\{h_k\}|H)$ is the Bayesian evidence for the data from $N$ observations $\{h_k\}$, for a model $H$, which is given by marginalizing over the hyperprior space
  \begin{equation}
  Z(\{h_k\}|H) = \int d\Lambda \, \mathcal{L}(\{h_k\}|\Lambda,H) \, \pi(\Lambda|H) .
  \label{eq:Z}
  \end{equation}
  From our (hyper)posterior distribution $p(\Lambda|\{h_k\})$, we construct confidence intervals for our hyperparameters.

  The odds ratio of two models is:
  \begin{equation}
  \mathcal{O}^i_j = \frac{Z(\{h_k\}|H_i)p(H_i)}{Z(\{h_k\}|H_j)p(H_j)} .
  \label{eq:O}
  \end{equation}
  We use the odds ratio to select between different models.
  Here, the $p(H_i)$ are the prior probabilities assigned to each model.
  In our study, we assign equal probabilities to each model.
  Thus, the odds ratio is equivalent to the Bayes factor:
  \begin{equation}
  B^i_j=\frac{Z(\{h_k\}|H_i)}{Z(\{h_k\}|H_j)}.
  \label{eq:bayes}
  \end{equation}
  We impose a somewhat arbitrary, but commonly used threshold of $|\ln(B)| > 8$ ($\sim 3.6 \sigma$) to define the point at which one model is significantly preferred over another.

  Now that we have derived a number of statistical tools, it is worthwhile to pause and consider what astrophysical questions we can answer with them.
  \begin{enumerate}[i.]
  \item If $p(\sigma_1, \sigma_2| \{h_k\})$ excludes $\sigma_1=\sigma_2=\infty$, then it necessarily follows that $p(\xi|\{h_k\})$ excludes $\xi=0$, and we may infer that at least {\em some} binaries merge through fieldlike models.
  \item If $p(\xi|\{h_k\})$ excludes $\xi=1$, we may infer that not all binaries can be formed via fieldlike models.
  \item If both $\xi=0$ and $\xi=1$ are excluded, then we may infer the existence of at least two distinct populations.
  \item If the $(\sigma_1,\sigma_2)$ posterior distribution  $p(\sigma_1, \sigma_2| \{h_k\})$ excludes $\sigma_1=\sigma_2=0$, we may infer that not all binaries are perfectly aligned.
\end{enumerate}
  In this way we can distinguish between different formation channels or specific models, i.e., perfect alignment in case (iv).

  We employ Bayes factors to compare our population models.
  We calculate evidences for three hypotheses:
  \begin{enumerate}[i.]
  \itemsep0em
  \item $Z_\text{dyn}$ -- Dynamic formation only, $\xi=0$.
  \item $Z_\text{field}$ -- Field formation only, $\xi=1$.
  \item $Z_\text{mix}$ -- Mixture of field and dynamic, $\xi\in[0,1]$.
  \end{enumerate}
  We then define three Bayes' factors to compare these three hypotheses:
  \begin{enumerate}[i.]
  \itemsep0em
  \item $B^\text{mix}_\text{field} = Z_\text{mix}/Z_\text{field}$.
  \item $B^\text{mix}_\text{dyn} = Z_\text{mix}/Z_\text{dyn}$.
  \item $B^\text{field}_\text{dyn} = Z_\text{field}/Z_\text{dyn}$.
  \end{enumerate}
  In the next section, we apply these tools to a variety of simulated data sets in order to show under what circumstances we can measure various hyperparameters and carry out model selection.

\section{Simulated population study}\label{MC}
  We use a simulated population to test our models.
  For the sake of simplicity, we construct a somewhat contrived population in which every binary shares some parameters corresponding to the best-fit parameters of GW150914:
  \begin{itemize}
  \item $(m_1, m_2)=(35 M_{\odot},30 M_{\odot})$.
  \item $d_L=\unit[410]{Mpc}$.
  \item $(a_1, a_2)=(0.6, 0.6)$.
  \end{itemize}
  Here, $d_L$ is luminosity distance and $(a_1, a_2)$ are the black hole spin magnitudes.
  The remaining extrinsic parameters (sky position and source orientation) are sampled from isotropic distributions.
  We emphasize that the distance and mass and spin magnitude distributions are not representative of the full population of black hole binaries, which is poorly constrained.
  These distributions represent a subset of GW150914-like events, chosen for illustrative purposes.
  In reality, for every GW150914-like event, there are likely to be a large number of more distant (and possibly lower mass) events, which contribute relatively less information about spin.

\begin{figure}
  \includegraphics[width=\linewidth]{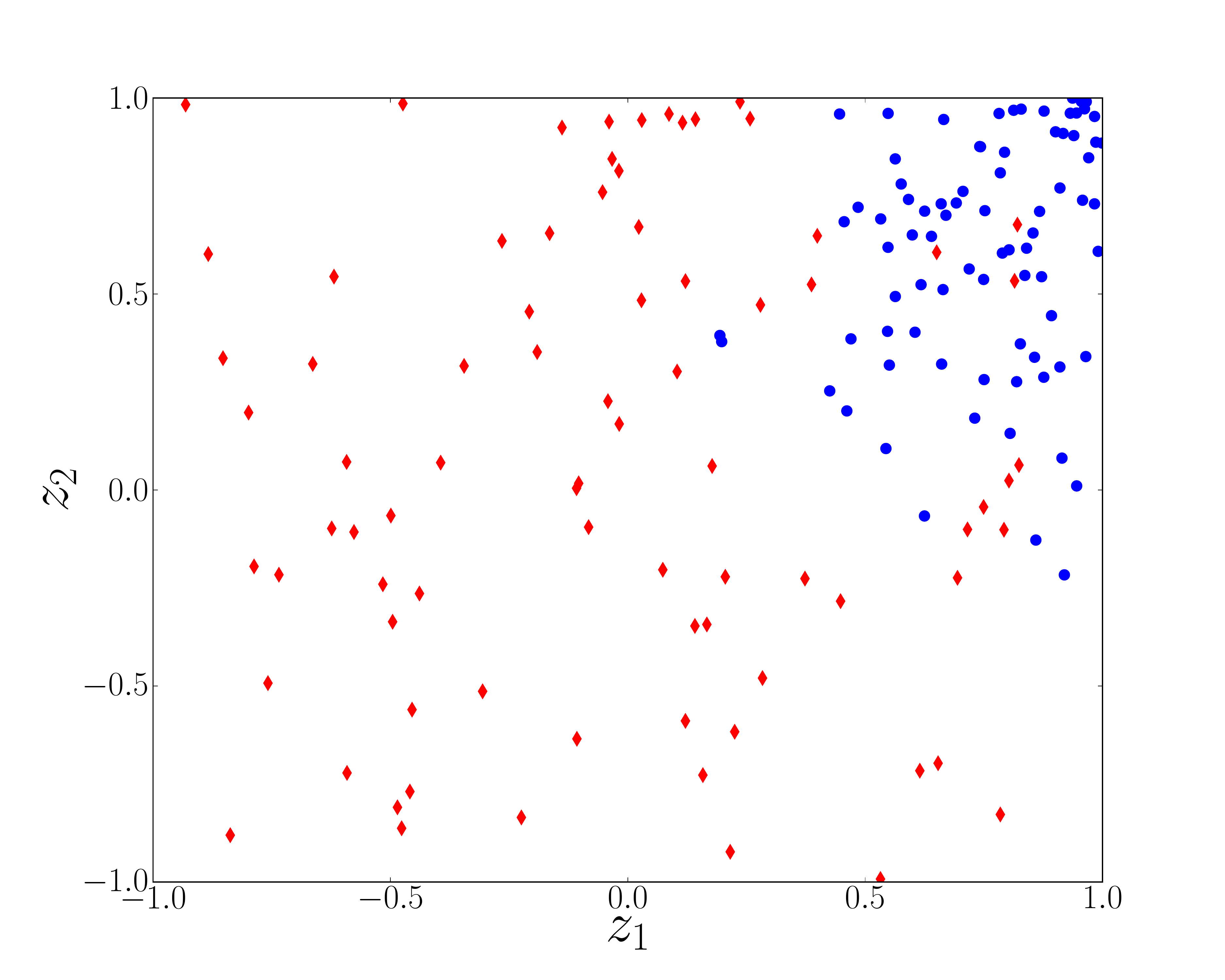}
  \caption{Simulated spin misalignment parameters $(z_1, z_2)$ for the different populations of binary black holes used in our study.
  Red diamonds are drawn from the isotropic distribution $p_0$ while the blue circles are drawn from the aligned distribution, $p_1(z_1, z_2|\sigma_1,\sigma_2=0.3,0.5)$; see Eq.~(\ref{eq:ps}).}
  \label{injections}
  \end{figure}

  We inject 160 binary merger signals into simulated Gaussian noise corresponding to Advanced LIGO at design sensitivity \cite{aLIGO, Abbott2016}.
  Of these, we generate 80 distributed according to $p_0$ and 80 distributed according to $p_1$; see Eq.~(\ref{eq:ps}).
  The injected values of $(z_1, z_2)$ are shown in Fig.~\ref{injections}.
  The red diamonds correspond to the $p_0$ dynamical model and the blue circles to the $p_1$ fieldlike model.
  From these we construct ``universes" summarized in Table \ref{table:universes}.
  Each universe contains a different mixture of field and dynamical binaries.
  In every universe, $(\sigma_1,\sigma_2)=(0.3, 0.5)$.

  \begin{table}[t]
  \begin{ruledtabular}
  \begin{tabular}{cccc}
  Universe & $\xi$ & $\sigma_1$ & $\sigma_2$ \\ \hline
  A & 0   & N/A & N/A \\
  B & 0.1 & 0.3 & 0.5 \\
  C & 0.5 & 0.3 & 0.5 \\
  D & 0.9 & 0.3 & 0.5 \\
  E & 1   & 0.3 & 0.5
  \end{tabular}
  \end{ruledtabular}
  \caption{Hyperparameters describing  different simulated universes.
  Here, $\xi$ is the proportion drawn from our aligned model and $(\sigma_1, \sigma_2)$ describe the typical misalignment angle; see Eq.~(\ref{eq:ps}).}
  \label{table:universes}
  \end{table}

\begin{figure*}[p]
  \centering
  \includegraphics[width=\linewidth]{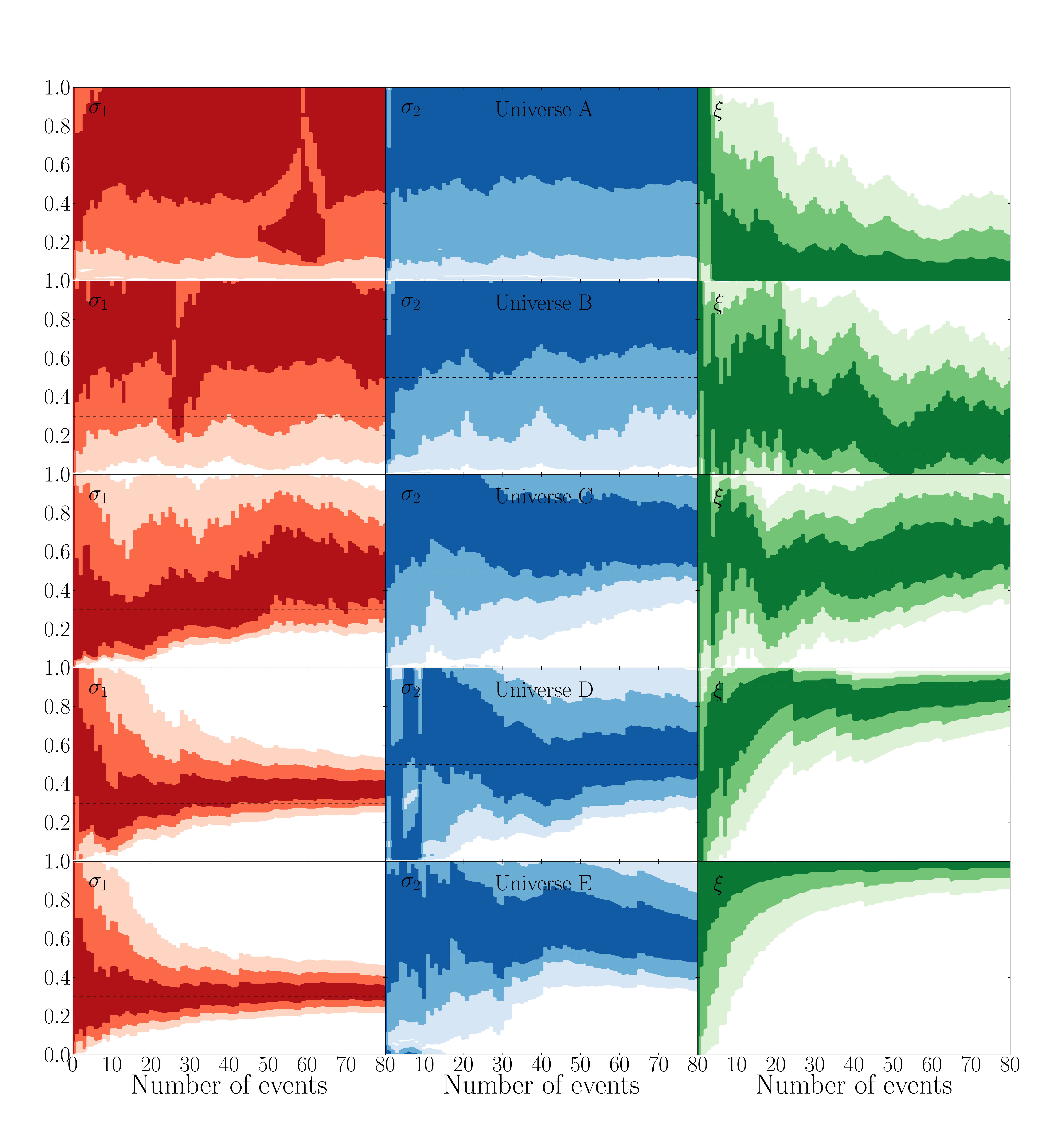}
  \caption{
  In each panel we plot $1\sigma$ (dark shading), $2\sigma$ (medium shading), and $3\sigma$ (light shading) confidence for different hyperparameters as a function of the number of events $N$.
  Each column represents a different hyperparameter: $\sigma_1$ (left), $\sigma_2$ (middle), and $\xi$ (right).
  Each row represents a different universe; see Table~\ref{table:universes}.
  From top to bottom, the universes are A, B, C, D, and E.
  The dashed line indicates the true hyperparameter values.
  The highest likelihood values of the three parameters after 80 events are shown on each panel along with the width of the $1\sigma$ confidence interval.
  }
  \label{fig:conf}
  \end{figure*}

\begin{figure*}[p]
  \vspace{-2cm}
  \includegraphics[width=\linewidth]{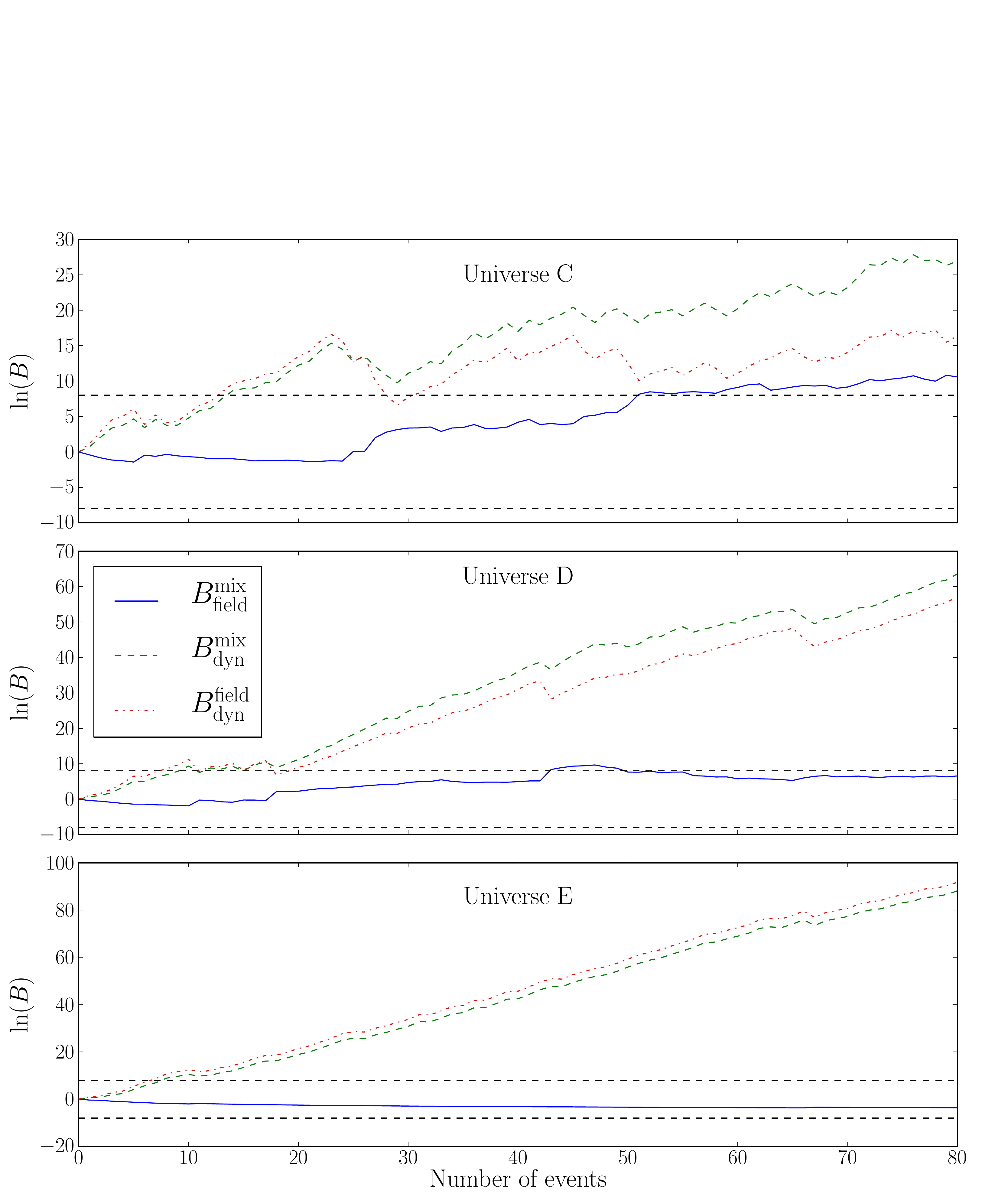}%
  \caption{
  Log Bayes factors as a function of the number of GW150914-like events.
  The dot-dashed red line shows $B_\text{dyn}^\text{field}$ comparing the pure-field hypothesis to the pure-dynamical hypothesis.
  The dashed green line shows $B_\text{dyn}^\text{mix}$ comparing the two-population hypothesis to the dynamical hypothesis.
  The solid blue line shows $B_\text{field}^\text{mix}$ comparing the two-population hypothesis to the pure-field hypothesis.
  The dashed lines denotes $|\ln(B)|=8$, our threshold for distinguishing between models.
  Each panel is a different universe.
  The top panel is universe C (equal mixture of field and dynamic).
  With $\lesssim40$ events, there is a strong preference for the two-component hypothesis over the pure-dynamic hypothesis.
  After $\sim50$ events there is a preference for the two-component hypothesis over the pure-field hypothesis.
  The center panel is universe D (majority field with some dynamic).
  With $\lesssim10$ events, there is a strong preference for the two-component and pure-field hypotheses over the pure-dynamic hypothesis.
  There is a preference for the correct two-population hypothesis over the pure-field hypothesis.
  The bottom panel is universe E (pure field).
  With $\lesssim10$ events, there is a strong preference for the two-component and field hypotheses over the dynamic hypothesis.
  There is a marginal preference for the correct field hypothesis over the two-population hypothesis.
  }
  \label{fig:bayes}
  \end{figure*}

  For each universe, we present the results of the methods described above.
  In Fig.~\ref{fig:conf}, we plot the $1\sigma$ (dark), $2\sigma$ (lighter), and $3\sigma$ (lightest) confidence regions as a function of the number of GW150914-like events.
  In Fig.~\ref{fig:bayes}, we plot the three Bayes factors defined in Eq.~(\ref{eq:bayes}) as a function of the number of GW150914-like events.
  Each row in Fig.~\ref{fig:conf} and panel in Fig.~\ref{fig:bayes} represents a different universe.

  First we consider universe A, consisting of only dynamically formed binaries, $\xi=0$; see the top row of Fig.~\ref{fig:conf}. 
  Since all binaries form dynamically in this universe, $\sigma$ is undefined.
  We see that after $O(1)$ event we rule out $\xi=1$ at $3\sigma$ (the hypothesis that all binaries form in the field).

  Next we consider universe E in which all events are drawn from the aligned model, $\xi=1$; see the bottom row of Fig.~\ref{fig:conf} and the bottom panel of Fig.~\ref{fig:bayes}.
  For this universe, $\sigma_1=0.3$, $\sigma_2=0.5$.
  We rule out $\xi=0$ (dynamical only) at $3\sigma$ after $O(1)$ event.
  The Bayes factors also rule out all binaries forming dynamically after $\lesssim10$ events.
  The threshold $|\ln(B)|=8$ is shown by the dashed line.
  After 80 events, the $1\sigma$ confidence intervals for $\sigma_1$ and $\sigma_2$ have shrunk to $\sim30\%$ and the $1\sigma$ confidence interval for $\xi$ has shrunk to $3\%$.
  The Bayes factor comparing the two-population hypothesis to the purely field hypothesis $B^\text{mix}_\text{field}$ (the blue line in the bottom panel of Fig. \ref{fig:bayes}) does not strongly favor field-only formation.

  Universes B, C and D are mixtures of the field and dynamical populations.
  Of these, B and D have only $10\%$ drawn from the subdominant population.
  We recover marginally weaker constraints than the corresponding single population universes.
  The hypothesis that all binaries form through the dominant mechanism is disfavored at $1\sigma$ after a few tens of events for universes B and D, establishing a weak preference for the presence of two distinct populations.
  For some realizations we can rule out both one component models after 80 events, however generally we see a subthreshold preference for the mixture model.
  This is unsurprising since each one-population model is a subset of our two-population model.
  For universe C, an equal mixture of events drawn from the field and dynamical populations.
  Both $\xi=0$ and $\xi=1$ are excluded at $3\sigma$ after tens of events establishing the presence of two distinct subpopulations.

  For all five universes, the presence of a perfectly aligned component ($\sigma=0$) is excluded after fewer than 20 events.
  For many realizations this number is $<5$.
  For universes B, C and D (consisting of a mixture of field and dynamical mergers), we can rule out the entire population forming from one of the two channels after 10--40 GW150914-like events.
  When there is a large contribution from the aligned model, we observe that the allowed region for $\sigma_1$ becomes small faster than the allowed region for $\sigma_2$.
  There are two effects, which explain this.
  First, the secondary black hole's spin has a less significant effect on the waveform~\cite{Vitale2014, Vitale2017b, Vitale2017c}.
  The spin orientation of the secondary is therefore less well constrained for each event.
  This translates to a larger uncertainty for $\sigma_2$ compared to $\sigma_1$.
  Second, the width of the distribution of spin tilts is broader for the secondary black holes.
  This broader distribution is intrinsically more difficult to resolve.

\section{Spin maps}\label{spinmaps}
\begin{figure*}
  \includegraphics[width=\linewidth]{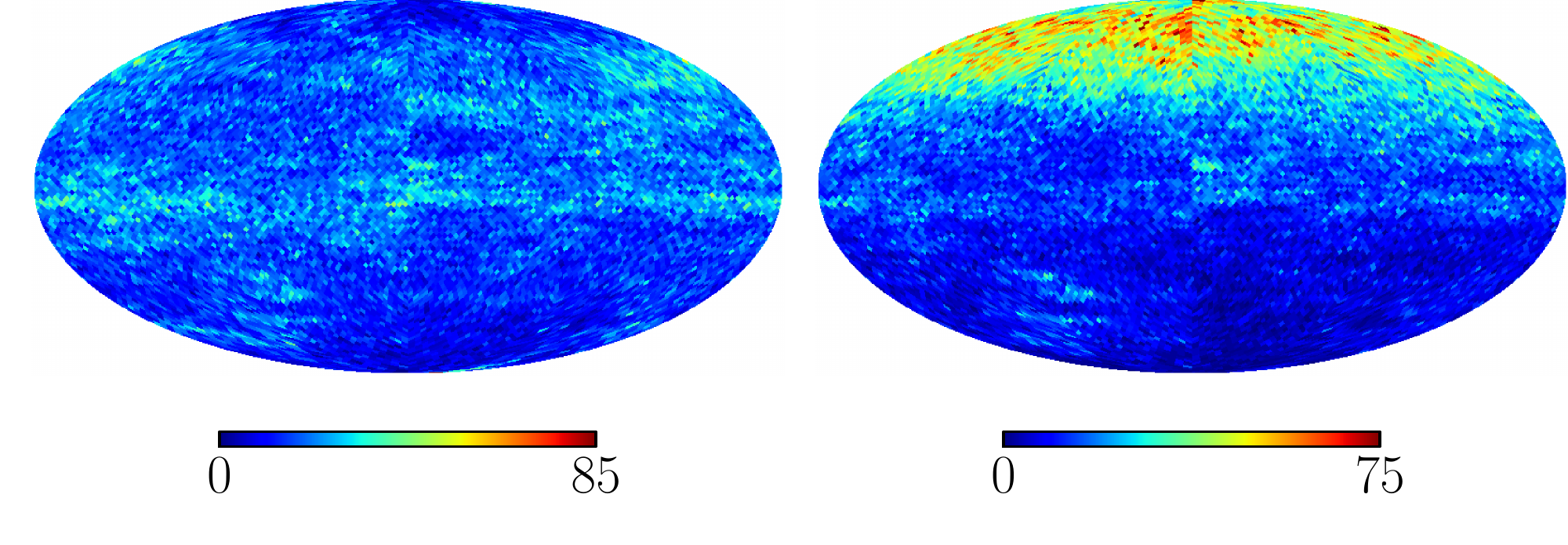}
  \caption{
  Spin maps: maps of posterior spin orientation probability density averaged over many realizations.
  The latitude is the spin tilt of the primary (more massive) black hole.
  The longitude is the angle between the projection of the black hole spins onto the orbital plane.
  The color bar is the number of posterior samples per $\unit[5]{\text{deg}^2}$ HEALPix bin.
  The left panel shows a spin map for 80 events drawn from universe A (see, Table \ref{table:universes}) in which every binary merges dynamically.
  The right panel shows 80 events drawn from universe C drawn in which, on average, 50\% of the events are drawn from the dynamical population while 50\% are drawn from the field population with $(\sigma_1,\sigma_2)=(0.3,0.5)$; see, Eq.~\ref{eq:ps}.
  The presence of a preferentially oriented population is seen as clustering around the north pole.
  }
  \label{maps}
  \end{figure*}

  In addition to our hierarchical analysis, we present a visualization tool for the distribution of spin orientations.
  We introduce ``spin maps": histograms of posterior spin orientation probability density, averaged over many events, and plotted using a Mollweide projection of the sphere defining the spin orientation, see Fig.~\ref{maps}.
  The maps use HEALPix~\cite{0004-637X-622-2-759}.
  For each posterior sample the latitude is the spin tilt of the primary black hole, $\theta_1$, and the longitude the difference in azimuthal angles of the two black holes, $\Delta\Phi$.
  The difference in azimuthal angles may give information about the history of the binary, specifically by identifying spin-orbit resonances at $\Delta\Phi=0,\pi$~\cite{Schnittman2004, Gerosa2013, Gerosa2014, Kesden2015, Gerosa2015, Trifiro2016}.
  These resonances, if detected, would appear as bands of constant longitude.
  We do not utilize azimuthal angle in this work and our injected distributions are isotropic in $\Delta\Phi$.
  In the future, it would also be interesting to produce ensemble spin disk plots (e.g., Fig. 5 of~\cite{PhysRevLett.116.241102}), showing the spin magnitude and orientation for a population of binaries.

  The spin maps in Fig.~\ref{maps} include contributions from 80 events for universes A and C (see Table \ref{table:universes}).
  This simple representation is useful because it provides qualitative insight into the distribution of spins and helps us to see trends and patterns that might not be obvious from our likelihood formalism.
  The north pole on these maps corresponds to spin aligned with the total angular momentum of the binary.
  We see the preference for the spin to be aligned with the angular momentum vector of the binary by the clustering in the northern hemisphere.

\section{Discussion}\label{discussion}
  The physics underlying the formation of black hole binaries is poorly constrained both theoretically and observationally.
  We do not know which of the proposed mechanisms is the main source of binary mergers: preferentially aligned mergers formed in the field versus randomly aligned mergers formed dynamically.
  We are also not confident in the predicted characteristics of binaries formed through either channel.
  We therefore create a simple (hyper)parametrization, describing the ensemble properties of black hole binaries.
  We demonstrate that we can measure hyperparameters describing the spin properties of an ensemble of black hole mergers with multiple populations.
  Previous work by Vitale \textit{et al.}~\cite{Vitale2017a} and Stevenson \textit{et al.}~\cite{Stevenson2017b} demonstrated that the fraction of binaries drawn from different populations can be inferred after O(10) events.
  We show that after a similar number of events, the shape of the spin-orientation distribution can be inferred using a simple hyperparametrization.
  We reproduce the finding from Stevenson \textit{et al.}, that $O(1)$ event is required to distinguish an isotropically oriented distribution, $\xi=0$, from a perfectly aligned distribution, $\xi=1$, $\sigma_1=\sigma_2=0$.
  After fewer than 40 GW150914-like events we can determine the properties of the dominant formation mechanism for all of our considered scenarios.
  We also introduce the concept of spin maps, which provide a tool for visualizing the distribution of spin orientations from an ensemble of detections.

  One limitation of our study is that, for the sake of simplicity, we employ a population of binaries with masses, distance, and spins fixed to values consistent with GW150914.
  The advantage of this simple model is that we are able to isolate the effect of spin orientation by holding other parameters fixed.
  The disadvantage is that the GW150914-like population is not a realistic description of nature.
  By changing from a population of binaries at a fixed distance to a population distributed uniformly in comoving volume, more events will be required for measurement of population hyperparameters.
  This is because most events, coming from the edge of the visible volume, will contribute only marginally to our knowledge of these hyperparameters.
  We assume fixed spin magnitudes of $a_1=a_2=0.6$.
  For a binary with aligned spins, this would imply $\chi_\text{eff}=0.6$.
  Based on recent LIGO detections, this might be optimistic.
  For GW151226, $\chi_\text{eff}=0.21^{+0.20}_{ - 0.10}$.
  For all other observed events, $\chi_\text{eff}$ is consistent with 0.
  This implies either that the observed black holes are not spinning rapidly or that the merging black holes observed so far possess significantly misaligned spins~\cite{PhysRevX.6.041015, GW170104, Farr2017}.
  If we have overestimated the typical black hole spin magnitude $a$, the number of events required to determine the distribution of spin orientation will increase.
  Implementing a theoretically motivated distribution of these parameters is left to future studies.
  Another area of future work is extending the method to other physically motivated spin orientation distributions.

\begin{acknowledgments}
    We thank Yuri Levin,  Simon Stevenson and Richard O'Shaughnessy for helpful comments.
    This is LIGO Document No.~DCC~P1700077.
    E.~T.~is supported through ARC FT150100281 and CE170100004.
\end{acknowledgments}

\bibliography{spin.bib}

\end{document}